# Attachment and antibiotic response of early-stage biofilms studied using resonant hyperspectral imaging

Yue Wang[1], Christopher P. Reardon[1], Nicholas Read[2], Stephen Thorpe[2], Adrian Evans[3], Neil Todd[3], Marjan Van Der Woude[4] and Thomas F. Krauss[1]

1. Department of Physics, University of York, Heslington, York, North Yorkshire, YO10 5DD, UK
2. Department of Biology, University of York, Heslington, York, North Yorkshire, YO10 5DD, UK
3. York Teaching Hospital NHS Foundation Trust, The York Hospital, York, North Yorkshire, YO31 8HE, UK
4. York Biomedical Research Institute, Hull York Medical School University of York, Heslington, York, North Yorkshire, YO10 5DD, UK

Corresponding author: Dr Yue Wang, Department of Physics, University of York, York YO10 5DD UK, tel: +44 1904322700, email: yue.wang@york.ac.uk

**Key words:** biofilm substratum, antimicrobial resistance, hyperspectral imaging, label-free sensing, biofilm detection

# Abstract

Many bacterial species readily develop biofilms that act as a protective matrix against external challenge, e.g. from antimicrobial treatment. Therefore, biofilms are often responsible for persistent and recurring infections. Established methods for studying biofilms are either destructive or they focus on the biofilm's surface. A non-destructive method that is sensitive to the underside of the biofilm is highly desirable, as it allows studying the penetration of antibiotics through the film. Here, we demonstrate that the high surface sensitivity of resonant hyperspectral imaging provides this capability. The method allows us to monitor the early stages of *Escherichia coli* biofilm formation, cell attachment and microcolony formation, in-situ and in real time. We study the response of the biofilm to a number of different antibiotics and verify our observations using confocal microscopy. Based on this ability to closely monitor the surface-bound cells, resonant hyperspectral imaging gives new insights into the antimicrobial resistance of biofilms.

# Introduction

Biofilms are commonly defined as communities of surface-attached microorganisms embedded in an extracellular matrix that are attached to a surface. In most natural environments, polymicrobial biofilms are predominant and their survival often relies on intercellular communication and interactions (1,2). In the last few decades, biofilms have attracted great attention in areas ranging from environmental studies to industrial water systems, from the food industry to chronic infections (3-6). In fact, 60-80% of human bacterial infections, including many bloodstream and urinary tract infections, are caused by bacterial biofilms (7-10). Biofilm-based infections are extremely difficult to cure, as biofilms are intrinsically much less susceptible to antimicrobial agents than non-adherent, planktonic cells. There are two main mechanisms for antimicrobial tolerance in biofilms. The first is the failure of an antimicrobial agent to diffuse and penetrate into the depth of the biofilm. This is due to the extracellular polymeric matrix that forms the biofilm and that is known to retard the diffusion of antimicrobial agents (11), and solutes in general. Hence the substratum, i.e. the surface-bound underside of the biofilm, remains protected. The second mechanism is that some of the biofilm cells experience nutrient limitation and therefore exist in a slow-growing or starved state (8-14). Because of their much slower metabolism, slow growing or non-growing cells are more tolerant of antimicrobial agents.

Recent studies have also shown that microorganisms growing in a biofilm are highly resistant to antimicrobial agents (15-19). The antibiotic resistance in biofilms is generally caused by mutations and is driven by the repeated exposure of the bacteria to high levels of antibiotics as a consequence of treating the biofilm-associated infections, for instance (20,21). The minimum inhibitory concentration (MIC), defined as the lowest concentration of an antimicrobial that will inhibit the visible growth of a microorganism, is often considered as the 'gold standard' for determining the susceptibility of microorganisms to antimicrobials (22,23). As a result from both biofilm tolerance and antimicrobial resistance, it is known that a subset of the sessile bacteria in biofilms can survive in the presence of up to 1000 times MIC, compared to the planktonic cells (24,25).

To account for this discrepancy, new pharmacodynamic parameters, such as the minimum biofilm inhibitory concentration (MBIC) and the minimum biofilm eradication concentration (MBEC) have been introduced. To evaluate antimicrobial activity on sessile bacteria, in-vitro systems with abiotic surfaces have been developed over the last decade (24-27). Closed systems (batch culture), such as multi-well plate assays, the Calgary device, and open systems (continuous culture), such as substratum suspending reactors and the flow cell systems, are some of the most used in-vitro biofilm models for susceptibility studies (28), but they often require staining and/or dissolving the biofilms for visualisation and quantification, so they are typically destructive and do not allow monitoring the progression of antimicrobial activity.

Despite all of the advances made in biofilm antimicrobial resistance and tolerance studies, a highly sensitive system that is able to monitor biofilm formation and dispersion in real time, non-destructively and quantitatively, is still missing. Monitoring the cell attachment and colonisation stage is paramount to this effort, as it allows for taking control of the early stages of biofilms and developing strategies for the prevention of mature biofilm formation. Similarly, monitoring the substratum of the biofilm is essential for understanding when antimicrobial action is completed. To address this challenge, we introduce the technique of resonant hyperspectral imaging to the monitoring of biofilms. We demonstrate the real time monitoring of cell attachment and the development of micro-colonies of *Escherichia coli* (*E.coli*) bacteria on the sensor surface with a range of initial inoculum density, starting from $2 \times 10^5$ CFU/ml. Due to the high surface sensitivity of the guided-mode resonance we employ, we are able to focus on the underside of the biofilm where antibiotic access is most challenging. We are able to provide a clear picture of the early stage of biofilm formation and of the response of the biofilm to a range of antibiotics with varying doses. The unique information provided by resonant imaging may support the development of novel therapies for treating biofilm-associated infections.

# Results

## Hyperspectral images of early-stage biofilms

The key element of our biofilm sensor is a grating which supports a guided-mode resonance (GMR) (29). The grating is fabricated in a 150 nm thick silicon nitride ($Si_3N_4$) film on a glass substrate. $Si_3N_4$ has a high refractive index to support the guided mode, it exhibits minimal absorption at visible and near-infrared wavelengths, and most importantly, it is biocompatible, chemically inert and mechanically robust (30,31). We first simulate the grating design using rigorous coupled-wave analysis (RCWA) and fabricate structures with a period of 569 nm and a filling factor (defined as the fraction of high-refractive index material in one period) of 80% for a resonance wavelength operating at 850 nm. The evanescent tail of the guided mode extends beyond the grating layer and into the sensing region and thereby defines the detection volume. The size of the detection volume is determined by the difference between the effective index of the GMR and the refractive index of the sensing region, as is apparent from fundamental guided-mode theory (32); the detection volume is typically 100 -200 nm in depth and only extends into the first bacterial layer of the biofilm. Therefore, the sensor is able to monitor the very early stages of biofilm formation and the biofilm substratum, and is not susceptible to the background signal that arises from the bulk biofilm and the surrounding media.

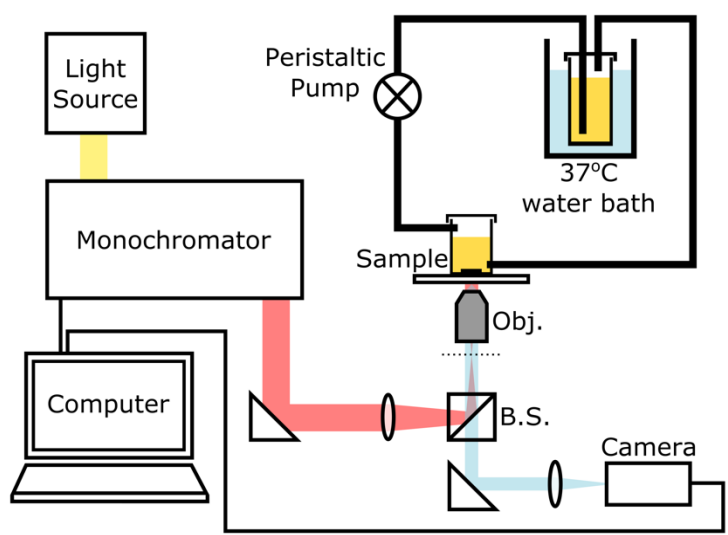

**Fig 1. Experimental setup.** A broadband light source (SM30, Leukos) combined with a grating-based monochromator is employed to generate a single illumination wavelength with a spectral width of 0.6 nm, which is then guided into an inverted microscope with a 5x objective lens (Olympus NeoDplan). A GMR grating is mounted on the bottom of a flow cell with PDMS; the flow of *E.coli* culture is generated by a peristaltic pump with sterile silicone tubes from a reservoir of culture in a 37°C water bath. The flow cell is used to avoid detection of the planktonic cells that happened to land on the grating but are not involved in biofilm formation. The camera used here is a CoolSnap Myo (Photometrics) with 970 x 730 pixels. Images are captured using LabView, and image analysis and curve fitting are performed using MATLAB.

Figure 1 is a schematic drawing of the experimental setup. The grating is placed into a flow cell to simulate the typical environment of interest, for instance a water pipe or the inside of the bladder, and to discourage settling of the bacteria. The flow is generated by a peristaltic pump with sterile silicone tubes. We used *E. coli* cultured in Luria-Bertani (LB) medium with an initial concentration of $2 \times 10^8$ CFU/ml and kept the bacteria in a 37 °C water bath and flowed them through the sensor with a flow rate of 5 ml/s. Images from the sensor are formed hyperspectrally by taking a sequence of bright-field images that are typically obtained in less than 3 minutes depending on the camera's integration time (usually set to be 1 s) and the wavelength resolution (usually set to be 0.2 nm), each at a different illumination wavelength achieved via a tunable monochromatic light source. For example, it takes just over 2 minutes for a scan of 25 nm wavelength range with 0.2 nm resolution, i.e. 125 images, and a hyperspectral image of the resonance is then generated within seconds. By analysing the intensity values of each bright-field image, the resonance wavelength for each pixel can be determined. By plotting the resonance wavelengths of all pixels in the array, we produce the hyperspectral images of the grating at resonance (for example in Fig. 2(a-d)). The resulting hyperspectral image contains the spectral response at every pixel within the microscope's field of view. The region of interest (i.e. the center of the GMR grating) consists of 200 x 200 pixels on the camera (corresponding to approximately 250 x 250 um$^2$ in size). We then combine the resonance wavelengths of all 200 x 200 pixels in each hyperspectral image as a histogram (see Fig. 2(e-h)), and by fitting a Gaussian curve to this histogram, we obtain a central resonance wavelength ($\lambda_n$) for each hyperspectral image. It is worth nothing here that the distribution in the resonant wavelengths at time 0, defined as $\lambda_0$, arises predominately from nanofabrication tolerances in a GMR grating. This uncertainty follows a normal distribution (33); hence, a Gaussian distribution is used to determine the central resonant wavelength and the resonance uniformity. Finally, the shift in the central resonance wavelength ($\Delta\lambda = \lambda_n - \lambda_i$, while $\lambda_i$ is defined as the base value of the resonance wavelength, averaged over the first 30 min after the flow of culture is started) caused by the refractive index change given by the bacterial attachment and the formation of micro-colonies, is plotted over time (see Fig. 2(i)). It appears that the shift in resonant wavelength, $\Delta\lambda$, saturates after approximately 5 h, which indicates that the first layer of the biofilm has been established; while it is understood that the biofilm continues to grow, the technique is not sensitive to further increases in thickness. We also note here that in the plane of the grating surface, the spatial resolution is 2 $\mu$m along the grating grooves and 6 $\mu$m in the direction perpendicular to the grating grooves (34).

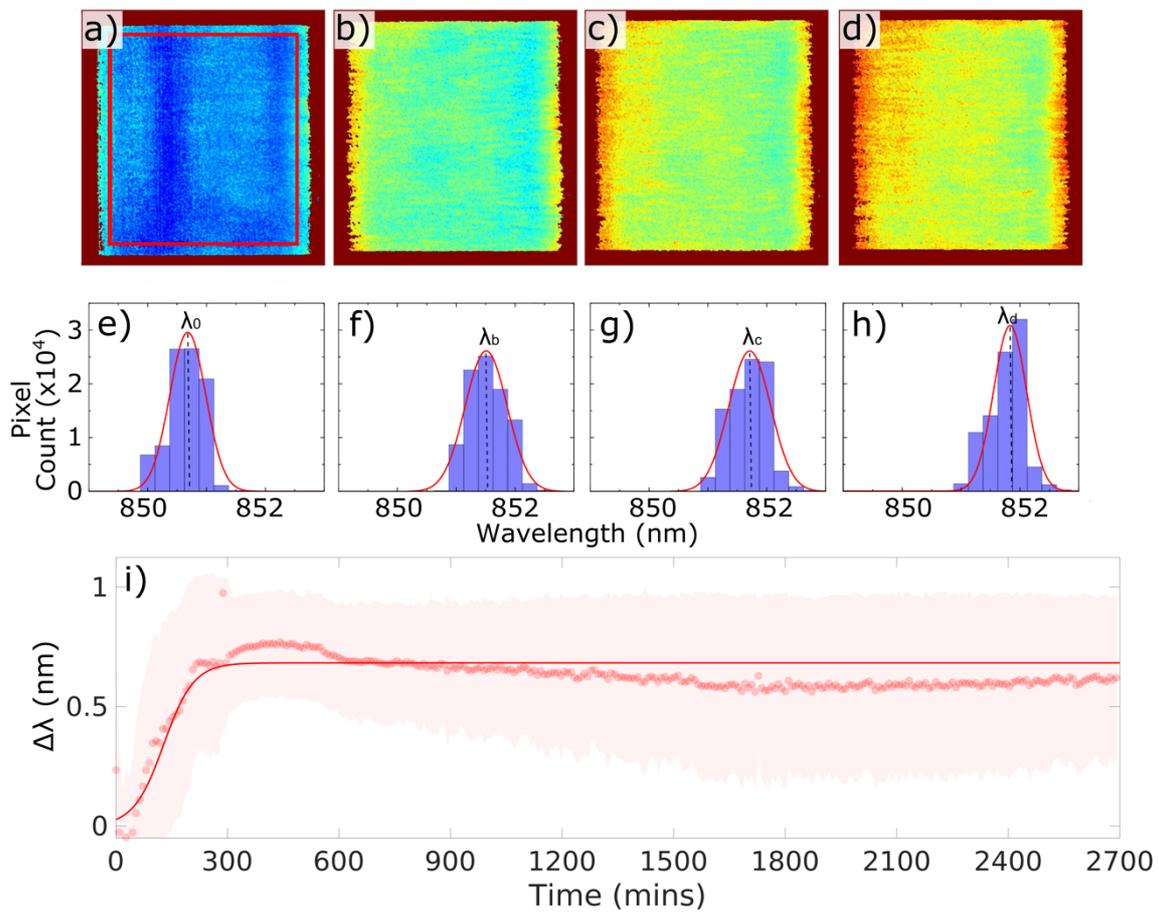

**Fig 2. Hyperspectral images and resonance wavelength shifts.** (a-d) Hyperspectral images generated by the biofilm sensor at times 0 h, 2 h, 3.5 h and 6 h from the start of the experiment. The colour represents the resonance wavelength; (e-h) histogram data of resonance wavelengths of all pixels in (a-d) correspondingly, and Gaussian fit (red curve) to determine the central wavelength: $\lambda_0$ (at 0 h), $\lambda_b$ (at 2 h), $\lambda_c$ (at 3.5 h) and $\lambda_d$ (at 6 h); (i) plot of the resonance wavelength shift, $\Delta\lambda$ ($\Delta\lambda = \lambda_n - \lambda_0$), against time (dots represent the central resonance wavelength $\lambda_n$ from each Gaussian fit; shaded areas illustrate standard deviations of corresponding Gaussian fits and delineates the distribution of the resonance wavelengths of all 200 x 200 pixels) and the solid line is a sigmoidal fit to the data. Data showing long term (45 hours) stability of the biofilm sensor with fresh LB media introduced into the flow cell at 24 h. The initial *E.coli* concentration used here is 2 x $10^8$ CFU/ml.

## Biofilm formation as a function of bacterial concentration

Cultures of *E.coli* were diluted to different concentrations, ranging from 2 x $10^5$ to 2 x $10^8$ CFU/ml. These dilutions were then used in separate experiments to inoculate the sensor and to study the rate of biofilm formation. We choose this range due to known clinical practice, where a specimen is considered positive for urinary tract infection in adults if the uropathogen concentration is greater than $10^5$ CFU/ml (35). We then set the sensor to obtain a hyperspectral resonant image at intervals of 9 minutes. Figure 3(b-e) shows that resonance wavelength shift is characteristic of early stage single species biofilm formation, which appears with a similar shape of typical growth curves of planktonic cells (36), although the mechanisms behind the curves are rather different. The biofilm formation profiles detected within a 15-hour window can be described by three phases. Initially, planktonic bacteria attach to the surface - although some attachments are reversible (named "planktonic" phase). The attached cells then aggregate to form micro-colonies while excreting extracellular polymeric substances, making their attachment irreversible (named "colonies" phase). The bottom-layer of the biofilm is then considered established and continues to mature into a multi-layered cluster ("mature" phase). The "mature" phase, which can be upwards of multiple micrometres in thickness, is beyond the detection volume of our sensor and is therefore not "seen", hence the resonance wavelength shift, $\Delta\lambda$, saturates. It is also known that the embedded biofilm cells do not divide and instead, their excess energy is used to maintain the extracellular matrix (11,37). This explains why the resonance wavelength, i.e. the refractive index on the sensor surface, remains constant after a certain period of time.

A modified sigmoidal function is used to model the biofilm formation curves, see Fig. 3. As discussed above, the characteristic "S"-shaped biofilm formation curve appears similar to the typical shape of planktonic bacterial growth curve, where sigmoidal models, such as Gompertz model, are commonly employed in analysis (36,38). Our modified sigmoidal function can be expressed in terms of the resonant wavelength shift, as follows:

$$\Delta\lambda = \frac{1}{1/A + exp[-\alpha * (t - t_{plank})]} \qquad \text{Eq. 1}$$

where $A$ is the amplitude of the profile, $t_{plank}$ is the time duration of the reversible cell attachment, which is indicated with a shaded region in Fig. 3, and $\alpha$ is the biofilm formation rate. The relationship between the bacterial seeding density and the "planktonic" phase time $t_{plank}$ is discussed in SI. Although it is intuitive that a higher initial inoculum density leads to a shorter biofilm formation time, the technique readily allows us to quantify this formation time. More importantly, the technique allows us to quantify the biofilm's response to antibiotics, as discussed in the next section.

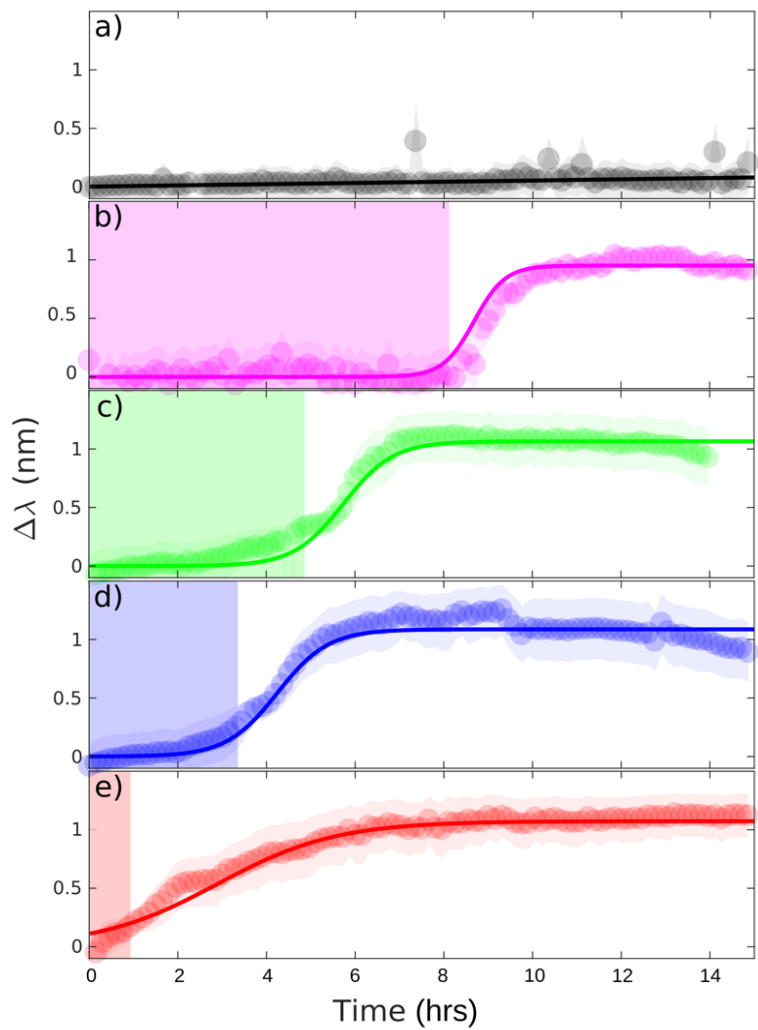

**Fig. 3. Biofilm formation time as a function of *E.coli* concentration.** (a) LB broth only, for reference. Concentrations of (b) 2 x 10$^5$, (c) 2 x 10$^6$, (d) 2 x 10$^7$, (e) 2 x 10$^8$ CFU/ml in LB. At least three independent replicates were performed for each concentration and reference. The solid lines are sigmoidal fits to the data and the fitting parameters are presented in Table 1. The shaded region indicates the duration of the reversible cell attachment, $t_{plank}$.

Table 1. List of fitting parameters in Figure 3, with the $R^2$ value for each fit.

| Initial concentration [cfu/ml] | $A$ [nm] | $t_{plank}$ [h] | $\alpha$ [nm/h] | $R^2$ |
|---|---|---|---|---|
| 2 x 10$^5$ | 0.952 | 8.22 | 0.428 | 0.92 |
| 2 x 10$^6$ | 1.07 | 4.93 | 0.267 | 0.94 |
| 2 x 10$^7$ | 1.09 | 3.40 | 0.260 | 0.96 |
| 2 x 10$^8$ | 1.07 | 0.94 | 0.112 | 0.97 |

## Response of *E.coli* biofilms to antibiotics

Next, we use the sensor to study the response of biofilms to antibiotic challenge, especially the response of the cells at the bottom layer of the biofilm. With monitoring the changes on the grating surface, we aim to detect which antibiotic can effectively disrupt the biofilm substratum. If the antibiotic causes a disruption to the distribution of the biofilm substratum, or partial or full detachment of the biofilm, we should observe a change in resonance wavelength of the grating. We use a concentration of 2 x 10$^8$ CFU/ml in order to speed up the process; using this concentration, the bottom layer of the biofilm is typically established within 5-6 hours (see Fig. S1). Our real-time, non-destructive monitoring capability then allows us to introduce antibiotics before, during and after a biofilm has been formed on the sensor surface.

For uncomplicated lower urinary tract infections in adults, current UK NICE (The National Institute for Health and Care Excellence) guidelines recommend nitrofurantoin (which inhibits the citric acid cycle as well as synthesis of DNA, RNA, and protein, also a bactericidal antibiotic) and trimethoprim (a folic acid synthesis inhibitor, a bacteriostatic antibiotic) as first line antibiotics due to their low risk of microbial resistance (39). We therefore choose these two drugs as our model antibiotics. The MIC and the minimum bactericidal concentration (MBC) were pre-assayed with planktonic *E.coli* cells in 96-well plates (see SI for details). The MIC of trimethoprim for this strain is determined to be <1 μg/ml and the MBC as 16-32 μg/ml. Trimethoprim with a dose of 1 μg/ml is therefore introduced into the culture at times 0 h, 3.5 h and 5.0 h. We make the following observations (Fig. 4):

- When the antibiotic is injected at time 0 h, it completely halts cell attachments, and no colonisation and subsequent multilayered biofilm formation is observed on the sensor surface;
- When the antibiotic is added at 3.5 h, the colonies stop developing immediately upon introduction of the drug, halting any further establishment of the biofilm;
- When 1x MIC of trimethoprim is added at 5.0 h (after the profile shows "saturation"), there is no observable change, i.e. the curve looks identical to the one without antibiotics added (see SI). This indicates that there is no change, i.e. in terms of colony density or detachment, at the bottom layer of the biofilm.

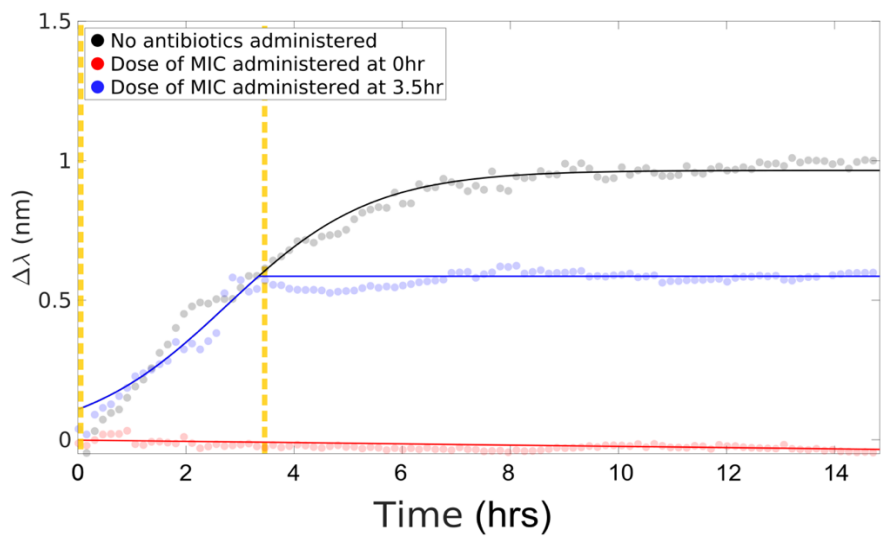

**Fig 4. Antimicrobial susceptibility testing in early-stage biofilms**. A 1x MIC dose of trimethoprim is introduced to the culture at 0 h and at 3.5 h (as indicated by the vertical dashed lines). The solid black line is a sigmoidal fit to the data. After introducing the antibiotic, the data are fit to a straight line.

In order to study the biofilm tolerance, we increase the dose of trimethoprim from 1 to 500x MIC (equivalent to 20x MBC approximately). The antibiotic solutions are administrated at 5 h from the start of the experiments, i.e. after the monolayer of biofilm is established. There is no significant effect on the detected resonance wavelength at any dose (see Fig. 5(a) for 500x MIC and more details in SI). Because the sensor probes the substratum of the biofilm, i.e. the volume that is in direct contact with the grating, the results indicate that even with extremely high doses of trimethoprim, the sensor does not detect significant changes. A separate study on nitrofurantoin reveals very similar results to trimethoprim (see SI for more information). We can therefore conclude that nitrofurantoin and trimethoprim are effective against planktonic cells, but not against biofilms formed by the same strain.

Confocal fluorescence microscope images confirm our findings (Fig. 5 b, c, g, h). Following the addition of 500x MIC of trimethoprim, introduced to the culture at 5 h, the biofilm is allowed to continue "growing" for at least a further 5 h with the antibiotic present. We observe that there is still a good bacterial coverage on the grating after the multiple wash steps that are required during the staining process, and the majority of cells attached to the surface are still alive (Fig. 5c, and more confocal fluorescence images in SI). This confirms that the bottom layer of the biofilm is undisrupted, in agreement with our sensor result. As clearly observed in Fig 5 (c), there are too many surviving bacteria in the 500x MIC case that remain attached to the sensor surface to consider this a biofilm eradication concentration (MBEC). This observation indicates that the MBEC is higher than 500x MIC, i.e. 0.5 mg/ml in this case, which is also consistent with other reports employing fluorescent testing methods (24,25). Colony counting on agar plates after 24 h incubation revealed that there are living cells in the remaining biofilms. So our recognition that trimethoprim is not effective against biofilms is confirmed by multiple control methods.

We then tested two other bactericidal antibiotics, i.e. rifampicin (an RNA synthesis inhibitor, can be bacteriostatic or bactericidal depending on concentrations) and ciprofloxacin (inhibits DNA replication and exhibits both bacteriostatic and bactericidal activities), and they have been suggested because of their efficacy against biofilms. Identifying the best antimicrobials against biofilms is an active area of research (13,40,41) that may benefit from a simple and real-time, non-destructive technique such as ours.

With our biofilm sensor, the disruption at the bottom of biofilm is observed as a shift of the resonance wavelength shortly after the administration of ciprofloxacin to a biofilm after 5 h of growth (Fig. 5(f)). Confocal microscopy with the live/dead assay confirms that the coverage of bacteria on the sensor surface is significantly reduced, and the cells in one remaining colony attached to the surface are also mostly dead (red colour) or injured with compromised cell membranes (yellow colour), shown in Fig. 5(h). This outcome is significantly different to what we observe with the other antibiotics, even though the phase contrast images of all these samples are rather similar (Fig. 5 d-e and i-j). The phase contrast images include not only the cells attached to the surface, but also the biofilms cells that are detached from the surface but near the surface as well as some planktonic cells. In contrast, confocal microscopy with the Live/Dead staining involves a few washing steps during the process (see Methods), therefore, the confocal images represent the biofilm cells that are still firmly attached to the sensor surface. In Fig. 5f, it certainly appears that 300x MIC ciprofloxacin results in a rapid and significant disruption on the bottom of the biofilm. Indeed, this disruption seems to cease after 5 hours and our hypothesis is that ciprofloxacin causes removal of majority of the biofilm on the sensor surface, leaving a small number of cells possibly with some exocellular matrix remaining on the surface, which is reflected by a small resonance wavelength shift after 10 h in Fig. 5(f). In short, our GMR sensor, compared to the conventional microscopic techniques, provides a non-destructive, rapid, dynamic sensing and analysis technique, which can provide real-time information of the very bottom of biofilm.

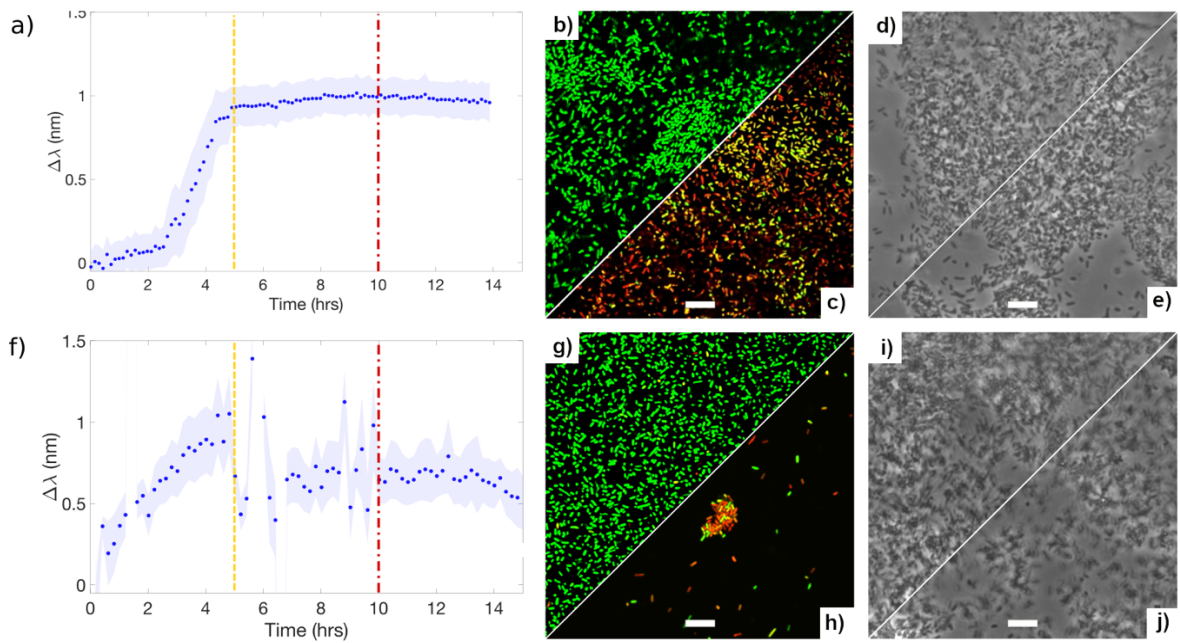

**Fig. 5. Three different techniques to monitor biofilms: GMR sensor (left), confocal fluorescence microscope (middle), and phase contrast microscope (right).** Trimethoprim (final dose of 500x MIC, a-e) and ciprofloxacin (final dose of 300x MIC, f-j) injected at 5 h (dashed line) from the beginning of the experiment and microscopic images are obtained after 10 h (dash dotted line) from the start of the experiments. Confocal microscopic images (b, c and g, h) and phase contrast images (d, e and i, j) are all taken at 10 h from the start of the

experiments; (c) and (h), (e) and (j) are confocal and phase contrast images of biofilms with antibiotic injected at 5 h from the start of the experiments, respectively; The scale bars in the micrographs are all 10 μm long.

For ease of comparison and reproducibility, all of these studies have been performed in LB broth and each set of experiments has been repeated for a minimum of three times. We have also performed total biofilm removal experiments with 70% ethanol wash followed by a piranha etch (a 3:1 mixture of sulfuric acid and 30% hydrogen peroxide). This protocol removes any organic residues on the grating surface including the extracellular matrix. A total recovery of the original resonance wavelength is observed, which indicates that complete removal of biofilms can be detected by our sensor. To ensure clinical relevance, we have also demonstrated that the technique is compatible with undiluted human urine (see SI).

## Discussion

In summary, we have introduced the technique of resonant hyperspectral imaging to monitoring the formation of *Escherichia coli* biofilms and have studied the response of these biofilms to a range of antibiotics. The key aspect that makes the technique so suitable for studying biofilms is that it only images the first few 100 nm from the bottom, i.e. the substratum, of the biofilm, which is the most important section of the film in terms of both early formation and antibiotic challenge. Because the technique is only sensitive to the substratum, the sensing signal is not affected by the media, the thickness of the biofilm or to any planktonic cells suspended in the media, so it is easy to achieve a high signal to noise ratio. Furthermore, within the same assay, a quantitative indication of the bacteria concentration can be readily obtained by monitoring the time it takes for the onset of biofilm formation. Given this real-time monitoring capability, the technique also allows antimicrobial resistance testing in-situ, therefore offering the opportunity of quantitatively analysing the action and progression of different antibiotics against the biofilm within minutes to a few hours. The non-destructive nature of the technique is a unique advantage, compared to, e.g., fluorescence-based microscopy techniques.

We envisage that our system can be miniaturised into a hand-held device that could be implemented as a diagnostic within a clinical environment; such a realization would allow for, e.g. the tailoring of antibiotic therapies to individual patients. Alternatively, the technique could be used for in-field biofilm detection, such as within water treatment plants or domestic water delivery pipes, food processing plants and equipment, heating, ventilation, and air-conditioning (HVAC) systems etc.. One can also envisage further improvements in terms of response time by accelerating bacterial attachment by using antibodies or carbohydrate coatings such as mannose. Undoubtedly, in a real biological or industrial system, the environment is far more complicated and polymicrobial biofilms tend to dominate. Although the polymicrobial nature of biofilms would not limit the performance of our sensor, additional sample or surface treatments may be used with our sensing system in order to improve specificity. For instance, pre-filtering or surface functionalisation could be employed in biofilm related bloodstream infection studies in order to gain some insight on the roles of each species in a biofilm community. Reports have also shown that the growth dynamics of a polymicrobial biofilm involve some extra phases (42), which can be further explored using our sensor system in future work.

Overall, we believe that our technique may be widely applicable both in industry and healthcare for fundamental studies and for applications that require the detection and monitoring of biofilms. With its high sensitivity to any changes in the substratum of biofilms, our sensor can be used in conjunction with other conventional techniques in order to gain a comprehensive understanding of biofilm formation, tolerance and antimicrobial resistance.

# Methods

## Bacterial strain and growth conditions

*E.coli* K-12 carrying the F conjugative plasmid, also known as *E.coli* TG1, is chosen in this work as it has been shown previously to promote thick, mature biofilms within a short period of time (43,44). A single colony is inoculated into a test tube containing 25 ml of liquid Luria-Bertani (LB) broth and placed in an orbital rotating shaker at 100 rpm overnight at 37°C. Some of this culture is then transferred and diluted within a fresh tube of LB broth to obtain the required concentrations of the microbial suspension. Optical density (OD) measurements are performed alongside separate cell counting on LB agar plates to determine bacterial concentrations. For example, $OD_{600nm}$ = 0.2 corresponds to a $2 \times 10^8$ CFU/ml concentration.

## Fabrication of resonance gratings and flow cell

In order to fabricate the gratings, a 150 nm $Si_3N_4$/glass substrate is cleaned in a piranha solution (hydrogen peroxide: sulfuric acid = 1:3 ratio), rinsed in acetone and isopropanol, and dried with nitrogen. The substrate is then coated with e-beam resist (ARP-13, AllResist GmbH), spun at 5000 rpm for 60 s, and baked at 180°C for 10 min, resulting in a film of approximately 400 nm. For charge dissipation during e-beam exposure, a thin film of aluminium (20 nm) is deposited on top of the resist using a thermal evaporator (HEX, Mantis). The grating pattern is defined in the e-beam resist using an electron beam lithography system (Raith GmbH Voyager 50 kV), followed by pattern transfer into the substrate using reactive ion etching with a gas mixture of $CHF_3$ and $O_2$. The depth of the grating in the substrate is 150 nm. The aluminium layer is removed in phosphoric acid and the residue of e-beam resist is removed with Microposit resist remover 1165 (MicroChem). The fabricated grating is then fixed to the bottom of a sterile quartz glass container. The flow of *E.coli* culture is generated by a peristaltic pump with sterile silicone tubes from a reservoir of culture, typically a total of 40 ml, in a 37°C water bath. Each set of experiments (including each *E.coli* concentration, references, each antibiotic and different concentrations) were repeated a minimum of 3 independent times.

## Live/dead staining and confocal microscopy

After a certain time of growth, the grating is taken out from the flow cell, gently washed with phosphate buffered saline (PBS) buffer (Sigma-Aldrich) three times to remove the remains of the medium and planktonic cells, as well as loosely attached non-biofilm cells, and then incubated for 15 min in the dark with the BacLight Live/Dead viability kit (Thermo Fisher Scientific) to stain the cells with two fluorescent dyes, SYTO® 9 (green) and propidium iodide (red). Samples were then rinsed twice with PBS buffer and sealed under a coverslip. The stained biofilms on the grating are visualised using a ZEISS LSM 880 Confocal Microscope with a plan apochromat 63x/1.4 oil-immersion objective. The Live/Dead viability kit stains live cells green, dead cells red and some cells appear yellow or orange indicating that those bacteria are injured with possibly compromised cell membranes (45).

### Antibiotic preparation and MIC assay

Stock solution of 10 mg/ml of four antibiotics, i.e. trimethoprim, nitrofurantoin, rifampicin and ciprofloxacin (Sigma-Aldrich), are prepared by dissolving in DMSO, diluting in deionized water and filter sterilizing (0.22 μm syringe filter, Fisher Scientific). A microdilution assay was also performed on 96-well plates by exposing *E. coli* TG1 to a serial dilution of trimethoprim, nitrofurantoin, rifampicin and ciprofloxacin, in order to determine the MIC of each drug. The reporter dye resazurin was added to indicate the viability status of the cells (46). For more details on the protocol, see SI. The MIC measurements were performed with OD = 0.2, i.e. $2\times10^8$ cfu/ml, and repeated for four times.

### Bacteria spiked urine sample

For the bacteria spiked urine sample shown in Fig. S8, an overnight-grown suspension of *E.coli* cells in LB broth is centrifuged at 8000 rpm for 10 min, and washed in syringe-filtered human urine (0.22 μm filter, Fisher Scientific). This process is repeated three times. Then the *E.coli* cells are resuspended in syringe-filtered human urine and the final concentration is made to be $2 \times 10^8$ CFU/ml in urine.

**Data and code availability**

The authors declare that all the data and MATLAB code supporting the findings of this study are available within the article and its supplementary information, or upon request from the corresponding author.

# Acknowledgement

The authors acknowledge financial support by the EPSRC of the UK (Grants EP/P02324X/1 and EP/P030017/1). Prof Thomas F Krauss acknowledges a Royal Society Wolfson Merit Award. Dr Yue Wang acknowledges a Research Fellowship awarded by the Royal Academy of Engineering. The authors also thank Dr Peter O'Toole, Joanne Marrison, Karen Hogg, Karen Hodgkinson, Graeme Park at the Bioscience Technology Facility at the University of York for their valuable assistance with the confocal microscopy, and Dr Donato Conteduca, Dr Giampaolo Pitruzzello for the thoughtful discussions.

**Competing interests**

The authors declare that there are no competing interests

**Author contributions**

YW, MVDW and TFK contributed to the conceptualization of the experiment; YW, CPR, NR and ST carried out the experiments; YW and CPR performed the data analysis; YW, AE, NT, MVDW and TFK contributed to the methodology and validation of the experiment; YW, CPR, NR, ST, MVDW and TFK wrote the manuscript.

# Supplementary information

I. Hyperspectral image vs. confocal microscope image

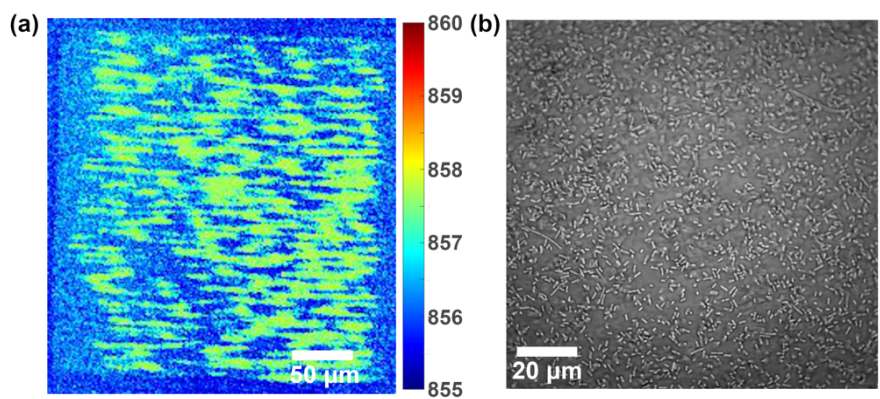

**Fig S1. (a) Hyperspectral image at 5h from the start of the experiment with initial e.coli concentration of $2 \times 10^8$ cfu/ml; (b) a corresponding confocal microscope image of the same biofilm.**

II. Uncertainty in resonance wavelength

The uncertainty in resonance wavelength, $\lambda_n$, is determined by monitoring the grating's resonance with a constant water flow. Over 7.5 h, the standard deviation ($\sigma$) of the resonance wavelength is measured to be 0.037 nm. And we estimate the smallest detectable shift to be $3\sigma$, which is 0.111 nm. The resonance wavelength shift, which is defined as $\Delta\lambda = \lambda_n - \lambda_0$, has a standard deviation $\sigma$ of 0.055 nm and $3\sigma$ of 0.157 nm.

III. Baclight Live/Dead viability assays

Samples are prepared in the flow cell for various lengths of time, i.e. 1 h, 3.5 h, 5 h and 10 h, with initial *E.coli* TG1 concentration of $2 \times 10^8$ CFU/ml. The gratings are then taken out from the flow cell, gently washed three times with phosphate buffered saline (PBS) buffer three times and stained with the Baclight Live/Dead viability assay. The washing steps are essential for removing loosely attached non-biofilm cells. The samples are then incubated at room temperature in the dark for 15 minutes. Fluorescent images are then taken with a ZEISS LSM 880 single-photon confocal microscope. It is worth noting here that for the short-time-grown sample (1 h in the flow cell), some bacteria were washed away from the grating surface during the staining process, due to weak attachment to the surface during this stage - this image therefore does not reflect the true nature of the sample in this case.

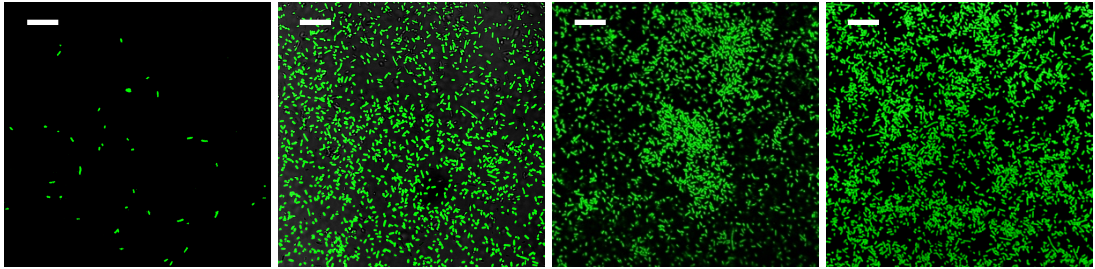

**Fig S2. Confocal microscopic images of early-stage biofilms on the sensor surface**. 1 h, 3.5 h, 5 h and 10 h biofilm samples are stained with the Live/Dead BacLight viability assay and examined with a ZEISS LSM 800 confocal microscope. The scale bars in the micrographs are all 10 μm long.

## IV. Biofilm formation time

The relationship between the bacterial seeding density and the "planktonic" phase time $t_{plank}$ is plotted in Fig. S3. Although it is intuitive that a lower initial inoculum density leads to a longer biofilm formation time, the technique readily allows us to quantify this formation time.

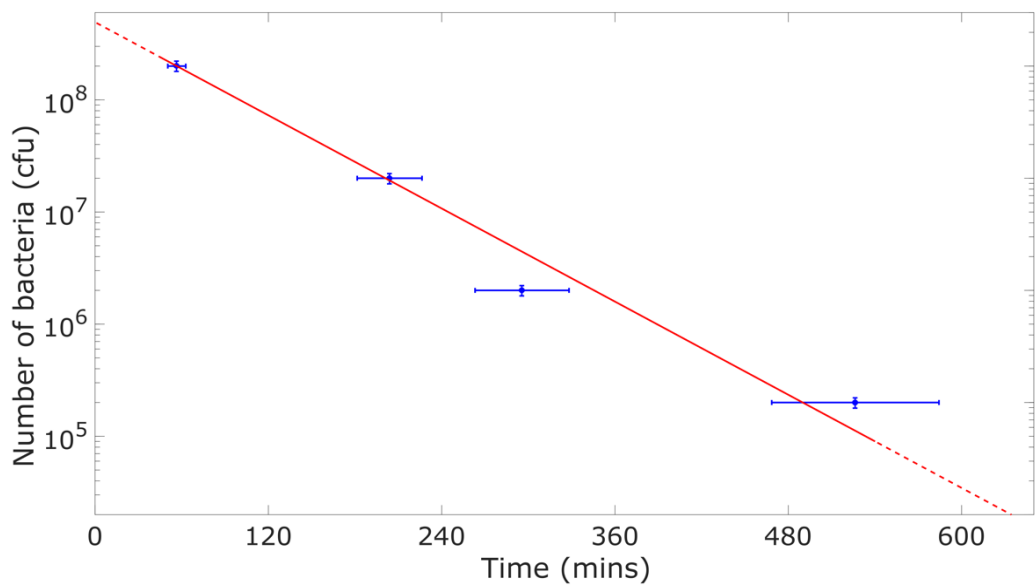

**Fig S3. Estimation of initial bacterial numbers.** Relationship between the initial bacterial concentration and the "planktonic" phase time duration of the early-stage biofilm ($t_{plank}$ in Eq.1) in a log-linear plot. The solid line is a linear fit to the data points. The error bars represent the standard deviations of repeated experiments.

## V. Determining the MICs

We use a resazurin-based microdilution assay to determine the minimum inhibitory concentrations (MICs) of different drugs. Trimethoprim, nitrofurantoin, rifampicin and ciprofloxacin are diluted from individual stock solutions to 1024 μg/ml. In a 96-well plate, column 1 of is used as a no antibiotic control and column 12 contains only LB broth as a sterility control (see Fig. S4). For the antibiotic serial dilution, 50 μl of the 1024 μg/ml antibiotic solution is first added to the 512 μg/ml column, and is well mixed with the liquid culture using a multichannel pipette before transferring to the next well, and repeated twice, to give a range of concentrations between 1 to 512 μg/ml, see Fig. S4.

*E. coli* TG1 are grown overnight in LB at 37°C and shaken at 200 rpm. Cells are washed three times in Mueller Hinton broth (MH) then re-suspended in MH before first being diluted to 0.1 OD to give standardised suspension of $10^8$ cfu/ml. The standardised suspension is diluted 1:100 then 50 μl added to each well to give a final bacteria concentration of $5.5 \times 10^5$ cfu/ml. 50 μl MHB is added to column 12 to give 100 μl of solution in all wells. After overnight incubation at 37°C, 30 μl of 0.015% solution of resazurin in MHB is introduced in each well and incubated for 4 hours. Results are validated by visual inspection of the wells.

The MIC microdilution experiments have been repeated for four times (twice with resazurin dye, twice without dye but the growth curves were monitored). The MICs of each drug are 0-1 μg/ml (trimethoprim), 4-8 μg/ml (nitrofurantoin), 8-16 μg/ml (rifampicin) and 0-1 μg/ml (ciprofloxacin). The MIC measurements were all performed with OD = 0.2, i.e. $2 \times 10^8$ cfu/ml diluted from overnight cultures.

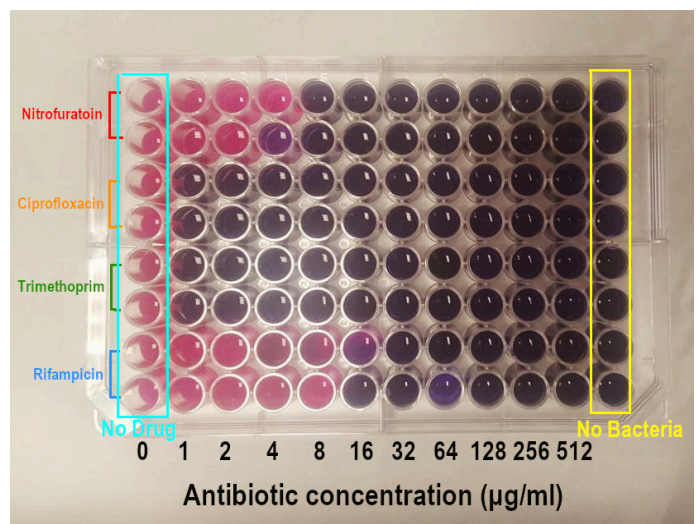

**Fig. S4. A resazurin microdilution assay with four different antibiotics.** The concentration of the four antibiotics, nitrofurantoin, ciprofloxacin, trimethoprim and rifampicin, is increased along the columns from 0 to 512 μg/ml. Each antibiotic has been tested in duplicate. The first column has no antibiotic and the last column has only LB broth with no bacteria. Pink colour indicates growth and dark purple means inhibition of growth.

## VI.   Antibiotics susceptibility test

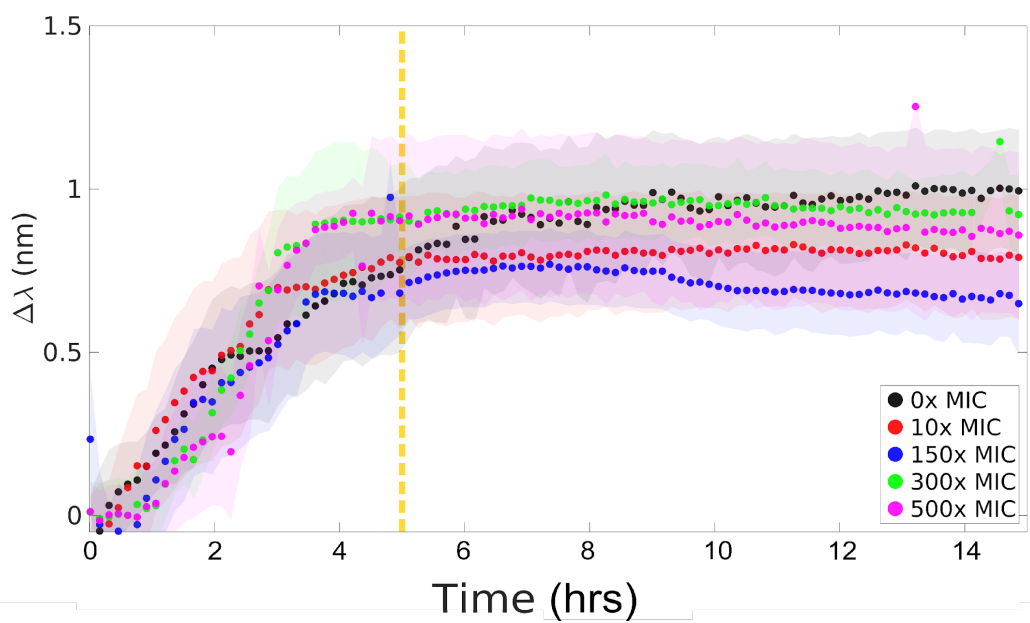

**Fig. S5. Trimethoprim susceptibility testing in established biofilms.** Trimethoprim at doses ranging from 10x to 500x MIC is introduced into the culture at 5 h (as indicated by the vertical dashed line). We observe no change within experimental accuracy, which suggests that trimethoprim has no effect on the biofilm.

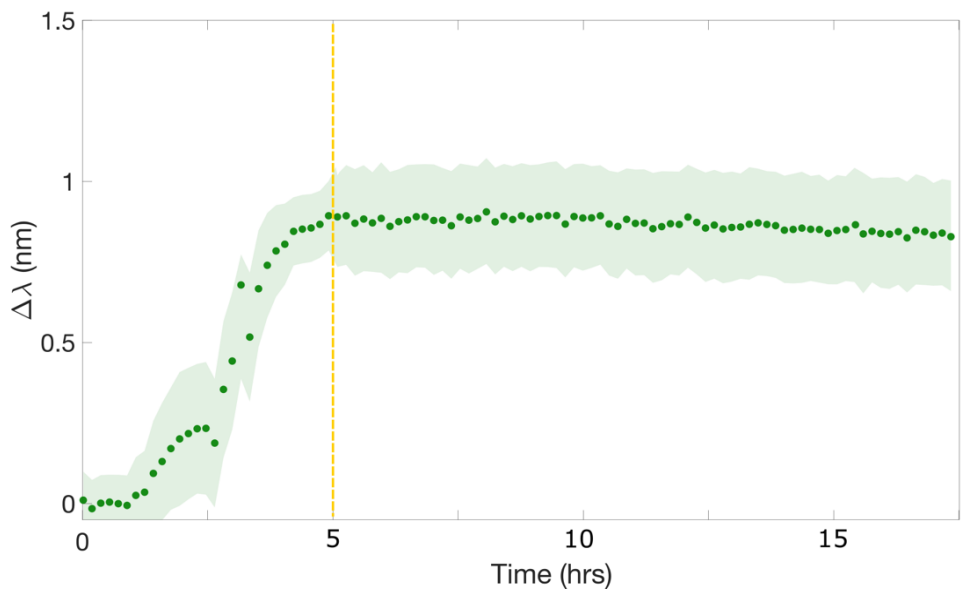

**Fig. S6. Nitrofurantoin susceptibility testing in established biofilm.** Nitrofurantoin with a concentration of 500x MIC is introduced into the culture at 5 h, and we observe no change in resonance wavelengths in the following 15 hours.

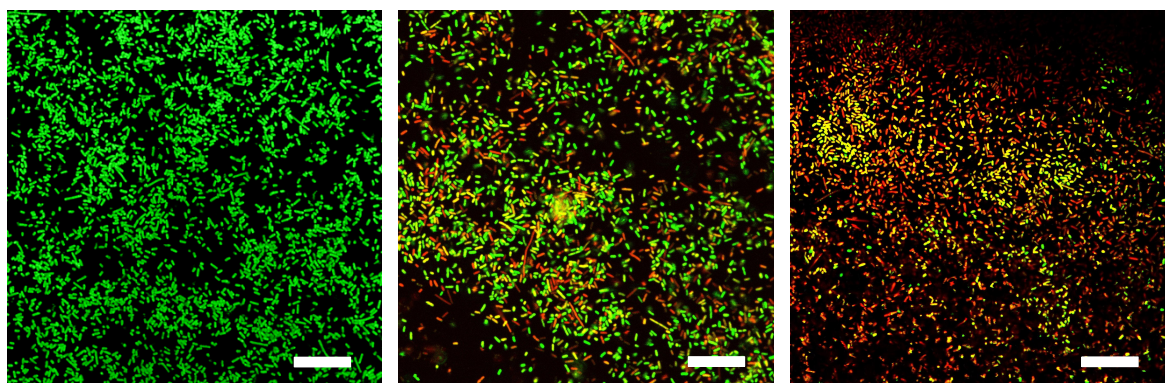

**Fig. S7. Confocal microscopic images of biofilms on the grating surface after 10 h from the start of the experiments**. With no antibiotic injected as a reference (left), with 150x MIC (middle), 500x MIC (right) of trimethoprim injected at 5 h from the start of the experiment. The biofilms were stained with the LIVE/DEAD BacLight viability stain assay at 10 h and cells with compromised membranes stain red, whereas cells with intact membranes stain green. The scale bars in the images are 10 μm.

## VII. Biofilm formation in bacteria spiked urine

Finally, we verify our technique's capability of working with a more complicated and biological growth culture, i.e. urine. A starting culture of *E.coli* TG1 was made from an overnight culture in LB broth. 50 ml cell suspension was spun down and washed three times in human urine (prefiltered using 0.2 μm pore size membrane filters). A suspension with *E.coli* inoculum density of 2 x $10^8$ CFU/ml, was flown through the sensor with the resonance wavelength shift monitored, see Fig. S8. Interestingly, the sensor detected full biofilm coverage within 60 minutes, i.e. significantly faster than in LB. This demonstrates that our sensor is capable of operating with a relevant biological matrix.

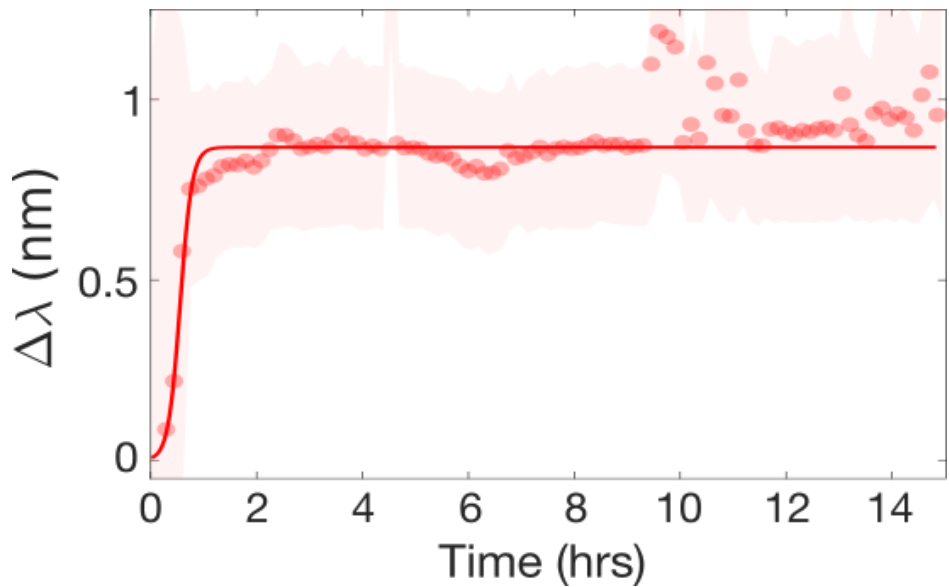

**Fig. S8. Biofilm detection in bacterial spiked human urine.** Biofilm growth is detected by the sensor in the flow cell with a suspension with *E.coli* inoculum concentration of 2 x $10^8$ CFU/ml in human urine.

26